\documentclass[%
 article,twocolumn,
superscriptaddress,
nofootinbib,
 amsmath,amssymb,
 aps,
 prd,
]{revtex4-1}

\newcommand{\parallelsum}{\mathbin{\!/\mkern-5mu/\!}}

\usepackage{graphicx}
\usepackage{dcolumn}
\usepackage{bm}
\usepackage{xcolor}
\usepackage{mathtools}
\usepackage{hyperref}
\usepackage[normalem]{ulem}

\newcommand{\AddrCFTP}{%
Departamento de F\'{\i}sica and CFTP, Instituto Superior T\'ecnico, Universidade de Lisboa, Av. Rovisco Pais 1, 1049-001 Lisboa, Portugal}

\newcommand{\AddrIEAP}{%
Institute of Experimental and Applied Physics, Czech Technical University in Prague,
Prague 160 00, Czech Republic}

\newcommand{\AddrOpava}{%
Research Centre for Theoretical Physics and Astrophysics, Institute of Physics,
Silesian University in Opava, CZ-74601 Opava, Czech Republic}

\begin{document}

\preprint{APS/123-QED}

\title{Neutrino-Induced Polarization Rotation in Active Galactic Nuclei Plasmas} 


\author{H.~B. C\^amara}\email{henrique.camara@cvut.cz}
\affiliation{\AddrCFTP} \affiliation{\AddrIEAP}

\author{A. Smetana}\email{adam.smetana@cvut.cz}
\affiliation{\AddrIEAP}

\author{A. A. Tursunov}\email{arman.tursunov@physics.slu.cz}
\affiliation{\AddrOpava}



\begin{abstract}
We study parity-violating birefringence induced by an asymmetric neutrino background in plasmas associated with active galactic nuclei (AGN). We derive a \emph{directionality factor} arising from the relative bulk motion between the neutrino medium and plasma, and show that it can produce an anomalous frequency dependence of the polarization-rotation angle, distinct from the $\omega^{-2}$ scaling of Faraday rotation. This anomalous scaling can occur either at the resonance plasma frequency condition $\omega \simeq \omega_p$, or when $E_\nu^{0}\simeq m_\nu \omega/\omega_p$ lies within the range of the neutrino energy spectrum. We estimate the effect for three scenarios: jets propagating through the cosmic neutrino background~(C$\nu$B), jets with an internal flux of high-energy neutrinos, and accretion-disk plasma permeated by the C$\nu$B. Of the three scenarios, the latter gives the largest rotation angle $\phi_{\rm d} \sim 10^{-35}\,\mathrm{rad}$, at X-ray frequencies. Although the predicted rotation angles are below current polarimetric sensitivity, the identified spectral signatures provide a theoretical framework for probing neutrino asymmetries and AGN plasma properties independent of magnetic field models.
\end{abstract}

\maketitle

\section{Introduction}
\label{sec:intro} 

The discovery of neutrino flavor oscillations~\cite{Kajita:2016cak,McDonald:2016ixn} implies massive neutrinos and lepton mixing, pointing to physics beyond the minimal Standard Model~(SM). Oscillation data~\cite{deSalas:2020pgw,Esteban:2024eli,Capozzi:2025wyn,Capozzi:2025ovi}, together with cosmological limits~\cite{Planck:2018vyg} and direct $\beta$-decay constraints from KATRIN~\cite{KATRIN:2024cdt}, indicate light neutrino masses of the order $m_\nu \sim 0.1$ eV. Despite their tiny masses and weak interactions, neutrinos are among the most abundant particles in the Universe as relics after the Big Bang, playing a crucial role in cosmology and astroparticle physics. Hence, complementary probes of neutrinos provide an unprecedented window onto the fundamental laws of Nature.

Although neutrinos are electrically neutral particles, they are subject to electromagnetic~(EM) interaction at the loop level in the SM -- most prominently they exhibit a magnetic moment predicted to be at most~$\sim10^{-19}\,\mu_B$. EM properties then translate into nontrivial optical properties of neutrino media. The dominant optical effect within the SM is the birefringence, or the optical activity that rotates the plane of linear polarization of the EM wave, caused by gyroelectromagnetic properties of the asymmetric neutrino chiral media~\cite{Royer:1968rg,Dicus:1993iy,Mohanty:1997mr,Abbasabadi:2001ps,Abbasabadi:2003uc,Karl:2004bt,Nieves:1988qz,Nieves:2005ya,Dvornikov:2013bca,Pal:2020bwo,Dvornikov:2020olb,Petropavlova:2022spq}. The birefringence of cosmic neutrino background~(C$\nu$B) intergalactic medium leads to a minuscule rotation of the linear polarization of the EM wave at the level of~$\sim10^{-40}\,\mathrm{rad}$, per the propagation length of Hubble radius~\cite{Dvornikov:2020olb}. It is relevant to search for various alternative scenarios where the effect could be enhanced like, e.g. in the attempt of one of us~\cite{Petropavlova:2022spq}. It has been demonstrated that the enhancement of the neutrino birefringence effect can be achieved by increasing the refractivity of the medium that approximately scales with electron density $n_e$. For example, propagating the optical wave in a silicon fiber provides an enhancement factor at the level~$\sim 10^{27}$ in comparison with the intergalactic plasma with $n_e\sim10^{-4}\,\mathrm{cm^{-3}}$.

Modern astrophysics has opened new pathways to explore the extreme environments around supermassive black holes~(SMBHs) in active galactic nuclei~(AGN). Namely, the jets in AGN provide extended structures of plasma over-densities of up to $\sim10^{4}\,\mathrm{cm^{-3}}$. Blazars, with their relativistic line-of-sight jets extending up to kiloparsec scales as observed in radio wavelengths, are subjects of a vital research, providing a sizable polarization rotation angle $\phi$ accumulated over the active path $l$ of the jet due to the Faraday rotation. The well-known dependence of the Faraday rotation measure~(RM) on the observed EM wave frequency $\omega$ is
\begin{equation}
    \frac{\phi}{l} = \frac{\phi_0}{l} + \frac{\text{RM}}{l} \ \omega^{-2} \ ,
    \label{eq:FRM}
\end{equation}
with $\phi_0 $ being a constant term and the RM for uniform plasma density and magnetic field is given by,
\begin{equation}
    \text{RM} = \frac{\omega_p^2 \omega_B l}{8\pi^2} \; , \; \omega_p = \sqrt{\frac{e^2n_e}{m_e}} \; , \; \omega_B = \frac{e B_{\parallelsum}}{m_e} \; ,
\end{equation}
where $\omega_p$ ($\omega_B$) is the plasma (cyclotron) frequency, $B_{\parallelsum}$ the parallel component of the magnetic field, and $m_e$ and $e$ are the electron mass and elementary charge, respectively. Note that, throughout this work, we use natural units unless explicitly stated otherwise. 

Faraday rotation has been measured directly at sub-millimeter radio wavelengths with modern polarimetric interferometry. In particular, the Atacama Large Millimeter/submillimeter Array~(ALMA) full-Stokes monitoring of Sagittarius~A* at the Galactic center yields a large, time-variable RM of the order of $\sim 10^{5}~{\rm rad\,m^{-2}}$~\cite{Wielgus:2023akf,EventHorizonTelescope:2024rju}. ALMA has likewise resolved Faraday rotation structure in nearby AGN jets, e.g. at the core of M87, for frequencies 85--101 GHz, the RM is of the order of $\sim 10^{4}~{\rm rad\,m^{-2}}$~\cite{Peng:2024dmb}. At sufficiently high photon energies, Faraday rotation rapidly becomes negligible due to its $\omega^{-2}$ scaling~\eqref{eq:FRM}. For example, in the X-ray band this suppression is extreme, even for the large RMs inferred in the immediate environments of black holes. The observed polarization angle is therefore expected to trace the intrinsic emission geometry. This expectation was borne out by recent Imaging X-ray Polarimetry Explorer~(IXPE) observations, which measure a stable polarization angle across the 2--8~keV band in accreting black hole systems and relativistic jets~\cite{galaxies12050054}. X-ray polarimetry thus provides a complementary, propagation-free probe of the innermost accretion flow and jet-launching regions. 

In this work, we estimate the polarization-rotation angle arising from the neutrino medium permeating the AGN jet and accretion disk, across the EM frequency range accessible to current polarimetric observations. Our primary goal is to draw attention to the directional dependence of the birefringence, first identified in~\cite{Petropavlova:2022spq}. This comes to effect only in cases where two components of the medium (neutrinos and plasma) have nonzero mutual bulk motion, as it is the case of either the AGN jet plasma that passes through the C$\nu$B, or is being passed through by ultra-relativistic neutrinos produced in the jet. We show that this \emph{directionality factor} can, under specific conditions, cause an anomalous $\omega$ dependence distinct from the Faraday rotation effect~\eqref{eq:FRM}. This paper is organized as follows. In Sec.~\ref{sec:birefringenceneutrinoplasma}, we derive the expressions for the birefringence angle induced by an asymmetric neutrino medium in a plasma, highlighting the role of the \emph{directionality factor}. Having established the theoretical framework, we analyze in Sec.~\ref{sec:jet} the polarization rotation induced by neutrino media in AGN jet plasma. In particular, we consider the contribution from the C$\nu$B in Sec.~\ref{sec:CnuB}, followed by ultra-high-energy~(UHE) neutrinos produced within the jet in Sec.~\ref{sec:UHEnu}. In Sec.~\ref{sec:accretiondisk}, we estimate the birefringence effect in the plasma of AGN accretion disks. Finally, our concluding remarks are presented in Sec.~\ref{sec:conclusion}.

\section{Birefringence from Asymmetric Neutrinos in Plasma}
\label{sec:birefringenceneutrinoplasma}

The rotary power from classical plane-wave framework is defined as (see e.g., Chapter XI of Ref.~\cite{LandauLifshitz1984})
\begin{equation}
    \frac{\phi}{l} \equiv \frac{1}{2}(|\mathbf{k}_-| - |\mathbf{k}_+|) \; ,
\end{equation}
where $\mathbf{k}_\pm$ are the wave vector eigenvalues of left and right-handed circular polarization eigenstates. By means of the wave equation for transverse modes, the presence of a medium is encoded in the transverse polarization function $\Pi_{\text{T}}$ and the parity-violating polarization function $\Pi_{\text{P}}$, which are components of the polarization tensor $\Pi_{\mu\nu}$ of the forward-scattered photon (see e.g., Chapter 6 of Ref.~\cite{Kapusta:2006pm}). The resulting rotary power is directly proportional to $\Pi_{\text{P}}$,
\begin{equation}\label{rotdef}
    \frac{\phi}{l}\simeq\frac{\Pi_{\text{P}}}{2\sqrt{\omega^2-\Pi_{\text{T}}}} = \frac{\Pi_{\text{P}}}{2\omega n}\,,
\end{equation}
being approximately valid for $\Pi_{\text{P}} \ll \omega^2 - \Pi_{\text{T}}$, with $n$ being the refraction index of the medium. 

It is a long-standing exercise to estimate the effect of a medium consisting of neutrino (anti-)particles with nonzero EM properties on the photon propagation~\cite{Royer:1968rg,Dicus:1993iy,Mohanty:1997mr,Abbasabadi:2001ps,Abbasabadi:2003uc,Karl:2004bt,Nieves:1988qz,Nieves:2005ya,Dvornikov:2013bca,Pal:2020bwo,Dvornikov:2020olb,Petropavlova:2022spq}. The leading effect in SM $\propto1/M^2_W$, where $M_W$ is the $W$-boson mass, is purely parity-violating being responsible for the birefringence. Namely, no matter how $\Pi_{\text{P}}$ is calculated, it is proportional to the neutrino-antineutrino density asymmetry $\propto(n_\nu-n_{\bar{\nu}})$ and the \emph{directionality factor} $K$ properly derived in~\cite{Petropavlova:2022spq},
\begin{equation}
	\Pi_{\text{P}}(k^2, K) = 2 C_\phi \; \frac{k^2}{m_e^2}(n_\nu - n_{\bar{\nu}}) \; K \; ,
\end{equation}
where the constant factor $C_\phi\propto G_F \, \alpha$, with $G_F$ being the Fermi constant and $\alpha$ the fine-structure constant. Note that $C_\phi$ comes from a loop calculation within the statistical quantum field theory of a chosen approximation scheme. The simplest approach, introduced in \cite{Mohanty:1997mr}, implements the effect of a plasma solely in the form of its refractivity, by replacing $k^2\rightarrow\omega_p^2$ in the one-loop vacuum expression, obtaining 
\begin{align}
     C_\phi^0 \equiv \frac{G_F \, \alpha}{3 \sqrt{2} \, \pi} \; .
\end{align}
In this work we use the improved estimate \cite{Dvornikov:2020olb} relying on two-loop implementation of the electron medium effect, leading to $\sim10^3$ enhancement,
\begin{align}
     C_\phi & = \frac{7 \pi}{2 \alpha} \times C_\phi^0  \; .
\end{align}

For the case of two independent background media, neutrinos and plasma, we have~\cite{Petropavlova:2022spq} 
\begin{equation}\label{eq:DirectionalFactor}
K=\omega\gamma|n-\beta\cos\theta_{ku}| \; , 
\end{equation}
where $\beta$ and $\gamma$ are the relativistic factors given by the bulk velocity of the neutrino medium with respect to the plasma, and $\theta_{ku}$ is the angle between the neutrino bulk velocity and the EM wave vector both in the rest frame of the plasma. In this work we take the simplified case of a non-relativistic and non-magnetized dilute electron plasma characterized by its plasma frequency
\begin{equation}
	k^2 = \omega_p^2 = \frac{4 \pi \alpha n_e}{m_e} \; ,
\end{equation}
and refractive index
\begin{equation}\label{eq:IndexOfRefraction}
	n(\omega,\omega_p) = \sqrt{ 1 - \frac{\omega_p^2}{\omega^2}} \; .
\end{equation}
For neutrinos of a given energy 
$E_\nu$, propagating through the plasma, the rotary power can then be written as
\begin{align} 
	\frac{\phi(E_\nu)}{l} &= C_\phi \ \frac{\omega_p^2}{m_e^2} \ \gamma(E_\nu) \ n_\nu(E_\nu) \left[1 - r_\nu(E_\nu)\right] \nonumber \\
	& \times \left|1 - \frac{\beta(E_\nu)}{n(\omega,\omega_p)} \cos \theta_{ku} \right| \; ,
	\label{eq:RotaryPowerAGN}
\end{align}
where we introduced the antineutrino-neutrino density ratio $r_\nu = n_{\bar{\nu}}/n_\nu$, taking values $r_\nu \geq 0$. In the above expression, we explicitly specify the $\omega$ and $E_\nu$ dependence of all quantities, and we emphasize that the rotation angle $\phi$ depends on the neutrino energy in the plasma. In addition, the neutrino density appears in the formula from averaging the statistical ensemble by performing the integration within the rest frame of the neutrino medium -- see Ref.~\cite{Mohanty:1997mr}. Furthermore, both $\gamma(E_\nu)$ and $n_\nu(E_\nu)$ depend on $E_\nu$ as a matter of the neutrino energy spectrum distribution. These quantities can be expressed in terms of an  energy-dependent differential neutrino flux $\Phi_\nu(E_\nu)$, leading to the differential rotary power
\begin{align} 
	\frac{1}{l} \frac{d \phi(E_\nu)}{d E_\nu} &= C_\phi \ \frac{\omega_p^2}{m_e^2} \ \Phi_\nu(E_\nu) \left[1 - r_\nu(E_\nu)\right] \nonumber \\
	& \times \left|1 - \frac{\beta(E_\nu)}{n(\omega,\omega_p)} \cos \theta_{ku} \right| \; .
	\label{eq:neutrinoflux}
\end{align}
The angle $\theta_{ku}$ is taken constant, and in what follows we consider only two values $\theta_{ku}=0,\pi$, neglecting possible divergences of the plasma, neutrino and EM wave fluxes.

\section{Effect of neutrinos in the Jet}
\label{sec:jet}

In this section, we analyze the polarization-rotation effect induced by an asymmetric neutrino medium within AGN jet plasma, first considering the C$\nu$B and then neutrinos produced inside the jet itself.

\subsection{Cosmic neutrino background}
\label{sec:CnuB}

C$\nu$B provides a universal neutrino medium through which all extragalactic EM radiation must propagate. Consisting of relic neutrinos from the early Universe, the C$\nu$B is expected to be nearly homogeneous and isotropic on cosmological scales. Polarized radiation emitted by relativistic AGN jets therefore experiences coherent forward scattering as it propagates through this neutrino background. The C$\nu$B is essentially at rest with respect to the Cosmic grid, while the jet plasma and EM wave move in the same direction towards us. Since the EM wave phase velocity is greater than the bulk velocity of the plasma $\beta$, we set $\theta_{ku} = \pi$ in Eq.~\eqref{eq:RotaryPowerAGN}.

The difference in number densities of cosmic neutrinos and antineutrinos for a given flavor $\alpha$, normalized to the photon density $n_\gamma \simeq 411~\mathrm{cm^{-3}}$, is~\cite{Iocco:2008va}:
\begin{align}
\eta_{\nu_\alpha} \equiv \frac{n_{\nu_\alpha} - n_{\bar{\nu}_\alpha}}{n_\gamma} & = \frac{\pi^2 \xi_\alpha}{12 \zeta(3)} \left( 1 + \frac{\xi_\alpha^2}{\pi^2} \right) \left(\frac{T_{\nu_\alpha}}{T_\gamma}\right)^3 \; ,
\label{eq:CnuBasymmetry}
\end{align}
written in terms of comoving neutrino chemical potential via the degeneracy parameters $\xi_\alpha \equiv \mu_{\nu_\alpha}/T_{\nu_\alpha}$, $\zeta(3) \simeq 1.20206$ and the SM value for the neutrino-photon temperature ratio is $T_{\nu_\alpha}/T_\gamma = (4/11)^{1/3}$. Recent analyses combining Cosmic Microwave Background data from Planck, BOSS and DESI Baryon Acoustic Oscillation measurements, large-scale structure observations, and primordial helium abundance determinations from EMPRESS, point to a small but nonzero flavor-equilibrated neutrino degeneracy parameter, $\xi_\nu \sim \mathcal{O}(0.05)$~\cite{Escudero:2022okz,Burns:2023sgx,Froustey:2024mgf,Li:2024gzf,Li:2025rjr}, corresponding to $n_\nu - n_{\bar{\nu}} \sim \mathcal{O}(5)~\mathrm{cm^{-3}}$~\footnote{Cosmic neutrinos undergo gravitational clustering, with their accumulation in galactic halos leading to a local overdensity of at most a factor of 10 relative to the cosmological average, see e.g. Refs.~\cite{Ringwald:2004np,deSalas:2017wtt,Zhang:2017ljh,Mertsch:2019qjv}. If taken into account that would lead to corresponding enhancement of the birefringence effect.}.

The C$\nu$B induced birefringence angle is given by, 
\begin{align} 
\frac{\phi}{l} &= C_\phi \ \frac{\omega_p^2}{m_e^2} \ \gamma \ \left|1 + \frac{\beta}{n} \right| \eta_{\nu_e} n_\gamma \; .
\label{eq:RotaryPowerAGNCnuB}
\end{align}
It is evident that the \emph{directionality factor} in~\eqref{eq:DirectionalFactor}, which encodes the $\omega$ dependence of the refractive index in~\eqref{eq:IndexOfRefraction}, gives rise to the above birefringence angle whose $\omega$ dependence is in principle distinct from that of Faraday rotation~\eqref{eq:FRM}. Nevertheless, in the regime $\omega_p/\omega \ll 1$, which applies to the AGN jet environment, the rotary power becomes approximately independent of the EM wave frequency, in agreement with Ref.~\cite{Dvornikov:2020olb}, and at second order in $\omega_p/\omega$ has the same $\omega$ dependence as~\eqref{eq:FRM}. Quantitatively, for the fiducial choice of AGN jet parameters, $n_e = 10^4~\mathrm{cm^{-3}}$, $l = 1~\mathrm{pc}$, $\beta = 0.9$, and a representative radio frequency $\omega = 1~\mathrm{GHz}$, we obtain the following estimate
\begin{equation}
    \phi^{\rm C\nu B}_{\text{j}} \sim 10^{-41}\,\mathrm{rad}\,.
    \label{eq:EstimateCnuB}
\end{equation}
This result is comparable to~\cite{Dvornikov:2020olb}, which examined the same effect in the intergalactic medium over a Hubble-scale path length $l \simeq 4.3$ Gpc, where the electron density is much lower, $n_e \simeq 10^{-4}~\mathrm{cm^{-3}}$. In contrast, AGN jets exhibit electron densities many orders of magnitude larger, even though their characteristic sizes are significantly smaller than the Hubble radius. 

Interestingly, we wish to point out that Eq.~\eqref{eq:RotaryPowerAGNCnuB} suggests that as the wave frequency approaches the local plasma frequency, $\omega \to \omega_p$ ($n \to 0$), the resonance can potentially occur, enhancing the neutrino-induced birefringence. However, in realistic AGN jets, the condition $\omega \simeq \omega_p$, typically marks a cutoff or turning point for wave propagation in an inhomogeneous plasma. Consequently, polarized radio emission that reaches the observer is unlikely to travel extended regions, and any enhancement of the C$\nu$B-induced birefringence remains strongly limited.

\subsection{Neutrinos produced in the jet}
\label{sec:UHEnu}

IceCube’s detection of an isotropic flux of astrophysical high-energy neutrinos, along with the association of individual events with extragalactic sources, has firmly established the feasibility of identifying UHE neutrino emitters~\cite{IceCube:2013cdw,IceCube:2013low,IceCube:2014stg}. Among the candidate sources identified by IceCube is the blazar TXS~0506+056~\cite{IceCube:2018dnn,IceCube:2018cha}, with other blazar candidates being studied in literature, e.g. PKS-class objects~\cite{Britzen:2021hfb,Padovani:2022wjk,VERITAS:2023eso}. Beyond IceCube, the large-volume detectors KM3NeT~\cite{KM3Net:2016zxf}/ARCA~\cite{KM3NeT:2018wnd,Muller:2023koj}, also targeting TeV--PeV neutrinos, are currently under deployment in the Mediterranean Sea. Notably, the KM3NeT Collaboration has recently reported the most energetic neutrino event to date with $E_\nu \sim 100$ PeV~\cite{KM3NeT:2025npi}, motivating studies of both a cosmogenic origin~\cite{KM3NeT:2025vut} and potential blazar associations~\cite{KM3NeT:2025lly}. Given this exciting experimental program, we aim to estimate the neutrino-induced birefringence effect from these UHE neutrinos produced in blazars~\footnote{We do not consider additional components of the jet-produced neutrino spectrum, namely the low energy neutrinos, as no observational data are presently available for them.}.

Relativistic jets in blazars can accelerate hadrons to UHE, producing TeV--PeV neutrinos, while simultaneously linking neutrino emission to the observed gamma-ray output~\cite{Hummer:2010vx,Biehl:2016psj,Fiorillo:2024jqz}. The dominant production channel is photohadronic $p\gamma$ interactions which arises from collisions of accelerated UHE protons with intense internal/external photon fields. The $p\gamma$ interactions then proceed via $\Delta^+$ resonance favoring charged pion $\pi^+$ production. The subsequent decay chain $\pi^+ \rightarrow \mu^+ \nu_\mu \rightarrow e^+ \nu_e \bar{\nu}_\mu \nu_\mu$, produces a UHE neutrino flux with an intrinsic neutrino--antineutrino asymmetry and a suppressed $\bar{\nu}_e$ component. Neutrinos then oscillate to an approximately flavor-equilibrated flux observed at Earth. Concrete modeling of photohadronic interactions in cosmic accelerators, such as in Ref.~\cite{Hummer:2010vx}, typically yields electron antineutrino-neutrino ratios of at most $r_\nu \sim 1/4$ for 100~TeV neutrinos, so that the electron neutrino-antineutrino asymmetry $(1-r_\nu) \sim 0.75$. However, the asymmetry depends on the photon spectrum and neutrino energy, which can noticeably impact the rotary power. A detailed modeling of these interactions is beyond the scope of this work. For the purpose of estimating the order of magnitude of the polarization rotation angle, we therefore adopt $(1-r_\nu) \approx 1$ in Eq.~\eqref{eq:neutrinoflux}. Furthermore, since the neutrinos are produced co-spatially with the observed polarized emission in the jet, the neutrino beam is naturally aligned with the photon propagation direction, we therefore set $\theta_{ku}\approx0$ in Eq.~\eqref{eq:neutrinoflux}. Moreover, it is clear that for the case under study the convenient quantity to use is the neutrino flux, which can be inferred from observations, instead of the local particle density, which is a derived quantity. The differential neutrino flux, over a given interval of neutrino energies $(E^{\mathrm{min}}_\nu,E^{\mathrm{max}}_\nu)$, can be modeled as a single power-law:
\begin{equation}
    \Phi_\nu = \Phi_0 \left( \frac{E_\nu}{E_0} \right)^{- \sigma} \theta(E_\nu-E^{\mathrm{min}}_\nu) \ \theta(E^{\mathrm{max}}_\nu-E_\nu) \; ,
    \label{eq:SPL}
\end{equation}
with $\sigma$ being the power-law spectral index, $\Phi_0$ the normalized differential neutrino flux and $E_0$ a reference neutrino energy. 

The overall birefringence effect is computed via
\begin{equation}
    \frac{\phi}{l} = C_\phi \frac{\omega_p^2}{m_e^2} \Phi_0 \int^{E^{\mathrm{max}}_\nu}_{E^{\mathrm{min}}_\nu} \mathrm{d}E_\nu \left( \frac{E_\nu}{E_0} \right)^{- \sigma} \left|1 - \frac{\beta(E_\nu)}{n}\right| \; .
    \label{eq:RotaryPowerAGNUHE}
\end{equation}
The integral above must be evaluated with care, since the absolute value develops a non-analytic behavior when the \emph{directionality factor} vanishes. This occurs when $\beta(E_\nu) = n(\omega,\omega_p)$, i.e. at the neutrino energy:
\begin{equation}
    E_\nu^0 = m_\nu\, \frac{\omega}{\omega_p} \; .
    \label{eq:Enu0}
\end{equation}
Moreover, because neutrinos are highly relativistic, $E_\nu \gg m_\nu$, implying $\gamma \gg 1$ and $\beta \simeq 1$. At the same time, the refractive index of the jet’s dilute plasma, at the photon frequencies currently probed by polarimetric observations, satisfies $\omega_p/\omega \ll 1$, the medium effectively has $n \simeq 1$. Consequently, to obtain analytical expressions we may safely perform a Taylor expansion of $\beta$ and $n$ in the small parameters $m_\nu^2/E_\nu^2$ and $\omega_p^2/\omega^2$, respectively, keeping only the leading-order terms. 

By integrating \eqref{eq:RotaryPowerAGNUHE}, below (or above) $E_\nu^0$, we obtain:
\begin{align}
    E^{\mathrm{max}}_\nu&<E_\nu^0 \ \ \ \Big(\mathrm{or}\ \ \ E^{\mathrm{min}}_\nu > E_\nu^0 \Big)\, : \nonumber \\ 
    \frac{\phi(\omega)}{l} &= \pm C_\phi \frac{\omega_p^2}{m_e^2} \Phi_0 \frac{m_\nu}{2} \nonumber \\ 
    &\cdot \Bigg\{ \frac{E_0^{\sigma}}{m_\nu} \frac{1}{1 + \sigma} \left[\frac{E_\nu^{\text{max}}}{\left(E_\nu^{\text{min}}\right)^{\sigma}} - \frac{E_\nu^{\text{min}}}{\left(E_\nu^{\text{max}}\right)^{\sigma}}\right] \frac{m_\nu^2}{E_\nu^{\text{max}} E_\nu^{\text{min}}} \nonumber \\
    &- \frac{E_0^{\sigma}}{m_\nu} \frac{1}{1 - \sigma} \left[\frac{E_\nu^{\text{max}}}{\left(E_\nu^{\text{max}}\right)^{\sigma}} - \frac{E_\nu^{\text{min}}}{\left(E_\nu^{\text{min}}\right)^{\sigma}}\right]  \frac{\omega_p^2}{\omega^2} \Bigg\} \; , \label{eq:belowabovephi}
\end{align}
written such that the terms within the brackets are dimensionless. The above result for the neutrino-induced birefringence angle exhibits the same dependence on $\omega$ as the one induced by Faraday rotation~\eqref{eq:FRM}, i.e. a dependence $\sim\mathrm{const.}+\omega^{-2}$. However, for the case where $E_\nu^0$ falls within $(E^{\mathrm{min}}_\nu,E^{\mathrm{max}}_\nu)$, the integration leads to:
\begin{subequations}
\begin{align}
    E^{\mathrm{min}}_\nu&< E_\nu^0 <E^{\mathrm{max}}_\nu\, : \nonumber \\ 
    \frac{\phi(\omega)}{l} &= C_\phi \frac{\omega_p^2}{m_e^2} \Phi_0 \frac{1}{2} m_\nu \nonumber \\ 
    &\cdot \Bigg\{ \frac{E_0^{\sigma}}{m_\nu} \frac{1}{1 + \sigma} \left[\frac{E_\nu^{\text{max}}}{\left(E_\nu^{\text{min}}\right)^{\sigma}} + \frac{E_\nu^{\text{min}}}{\left(E_\nu^{\text{max}}\right)^{\sigma}}\right] \frac{m_\nu^2}{E_\nu^{\text{max}} E_\nu^{\text{min}}} \label{eq:zerocte} \\
    &+ \frac{E_0^{\sigma}}{m_\nu} \frac{1}{1 - \sigma} \left[\frac{E_\nu^{\text{max}}}{\left(E_\nu^{\text{max}}\right)^{\sigma}} + \frac{E_\nu^{\text{min}}}{\left(E_\nu^{\text{min}}\right)^{\sigma}}\right]  \frac{\omega_p^2}{\omega^2} \label{eq:zero2} \\
    &- \frac{E_0^{\sigma}}{m_\nu} \frac{1}{1 - \sigma^2} \left(m_\nu\frac{\omega}{\omega_p} \right)^{-\sigma} 4 m_\nu \frac{\omega_p}{\omega} \Bigg\} \; , \label{eq:zeroA}
\end{align} 
\end{subequations}
where the \eqref{eq:zeroA} term exhibits a distinct anomalous $\omega$ dependence compared to the standard Faraday rotation~\eqref{eq:FRM}. This follows from the directional dependence, which is essential to account for, given that the two components of the medium -- neutrinos produced in the jet and the surrounding plasma -- exhibit nonzero relative bulk motion. As an example, for $\sigma \rightarrow 0$ (step function), where the differential neutrino flux is constant over $(E^{\mathrm{min}}_\nu,E^{\mathrm{max}}_\nu)$, the anomalous dependence is $\sim\omega^{-1}$, while for $\sigma \rightarrow 2$ (power-law) the anomalous dependence is $\sim\omega^{-3}$.

    \begin{figure}[t!]
        \centering
        \includegraphics[scale=0.6]{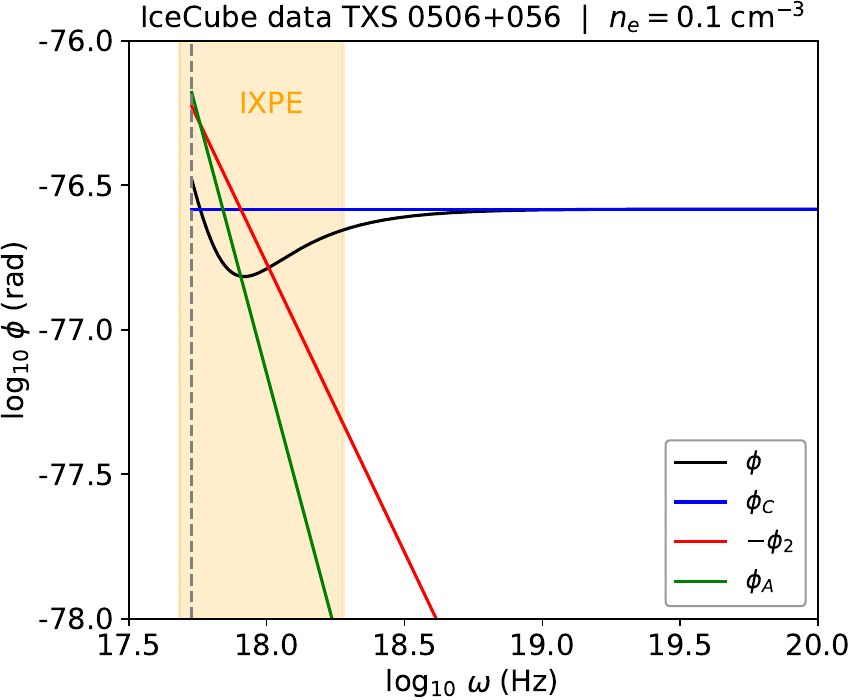}
        \caption{Birefringence angle $\phi$ in terms of frequency~$\omega$. Rotary power induced by neutrinos produced in the blazar source TXS 0506+056 identified by IceCube~\cite{IceCube:2024fxo}, for a plasma density $n_e = 0.1$ cm$^{-3}$. Total angle $\phi$ is given by the black curve, while its constant $\phi_C$ (quadratic $\phi_2$) [anomalous $\phi_A$] component, given by Eq.~\eqref{eq:zerocte} (Eq.~\eqref{eq:zero2}) [Eq.~\eqref{eq:zeroA}], is shown in blue (red) [green]. Orange shaded region corresponds to frequencies probed by IXPE~\cite{galaxies12050054}. To the left of the vertical dashed line $E_\nu^0$~\eqref{eq:Enu0} is no longer within $(E^{\mathrm{min}}_\nu,E^{\mathrm{max}}_\nu)$.}
    \label{fig:TXS_plot}
    \end{figure}
To  realistically estimate the magnitude of this effect we consider the blazar neutrino source TXS~0506+056 identified by IceCube. From the latest fit to the data, assuming a single power-law flux as in Eq.~\eqref{eq:SPL}, the collaboration determined a best-fit spectral index of
$\sigma= 2.58$ and per-flavor flux $\Phi_0 = 1.68 \times 10^{-18}~\text{GeV}^{-1}\,\text{cm}^{-2}\,\text{s}^{-1}\,\text{sr}^{-1}$, within an observed interval $(E^{\mathrm{min}}_\nu,E^{\mathrm{max}}_\nu) \in (3, 550)~\text{TeV}$, at $E_0 = 100~\text{TeV}$~\cite{IceCube:2024fxo}. By knowing that TXS~0506+056 is at a luminosity distance $D = 1.75~\text{Gpc}$ from Earth and adopting a jet length $l = 1~\text{pc}$, in what follows we use the simple geometric factor $(D/l)^2$ to estimate the local neutrino flux within the jet integrated over the full solid angle. Notice that, in order for $E_\nu^0$ to fall within the interval $(E^{\mathrm{min}}_\nu,E^{\mathrm{max}}_\nu) \in (3, 550)~\text{TeV}$, it is required a plasma-to-EM wave frequency ratio of $\omega_p/\omega \in (0.018, 3.33) \times 10^{-14}$. For an electron number density $n_e = 10^4\,\mathrm{cm^{-3}}$, this corresponds to frequencies in the range $\omega \in (1.69,\, 310) \times 10^{20}\,\mathrm{Hz}$. which are clearly far above those accessible to current polarimetric observations, being even above IXPE's sensitivity $\omega \in (0.483,\, 1.9)\,\mathrm{EHz}$~\cite{galaxies12050054}. However, as shown in Fig.~\ref{fig:TXS_plot}, for a plasma with lower density $n_e = 0.1$ cm$^{-3}$, the anomalous contribution $\phi_A$ of Eq.~\eqref{eq:zeroA} (green curve) lies within the IXPE frequencies (orange shaded region). However, for such EM wave frequencies and electron density, the overall value of the birefringence angle $\phi$ (black curve) is extremely small $\sim 10^{-77}$. For the case where $E_\nu^0$ falls below $(E^{\mathrm{min}}_\nu,E^{\mathrm{max}}_\nu)$ the overall effect is enhanced. Namely, using Eq.~\eqref{eq:belowabovephi}, with  $n_e = 10^4\,\mathrm{cm^{-3}}$, we obtain exceedingly small polarization-rotation angles of $\phi \sim 10^{-49} - 10^{-67}~\mathrm{rad}$, for typical radio-to-X-ray frequencies $\omega = \,1~\text{GHz} - 1~\text{EHz}$, respectively. 

For neutrinos produced in the jet, enhancing the overall birefringence beyond a larger plasma density requires higher neutrino fluxes. Assuming that, within a given energy interval, the flux reaches an optimistic upper limit such that the neutrino density is comparable to the local plasma density $n_\nu \sim n_e$, we can estimate the maximum expected rotation angle. For the same fiducial choices as the C$\nu$B case (see Sec.~\ref{sec:CnuB}), $n_e = 10^4~\mathrm{cm^{-3}}$, $l = 1~\mathrm{pc}$ and $\omega = 1~\mathrm{GHz}$, we obtain,
\begin{equation}
    \phi^{\rm UHE}_{\rm j} \sim 5 \times 10^{-44}\,\mathrm{rad}\,,
\end{equation}
being a few orders of magnitude smaller than the C$\nu$B estimate of Eq.~\eqref{eq:EstimateCnuB}. 

\section{Effect of neutrinos in the accretion disk}
\label{sec:accretiondisk}

In the previous section, we estimated the birefringence induced by neutrinos in the plasma environment of AGN jets. We now consider the complementary scenario of neutrinos in accretion disk plasmas, where the geometry, kinematics, and plasma conditions differ significantly from the jet. Notably, whether we consider the C$\nu$B permeating the disk or neutrinos produced locally within the accretion disk, the \emph{directionality factor}~\eqref{eq:DirectionalFactor}, relevant in the jet context, reduces here to unity. Unlike relativistic blazar jets, where polarized emission allows a well-defined jet plasma direction relative to the asymmetric chiral neutrino medium, for the accretion disk case there is no preferred bulk motion of neutrinos relative to the plasma. From Eq.~\eqref{eq:DirectionalFactor}, this corresponds to vanishing bulk velocity $\beta=0$ and $\gamma=1$, so that Eq.~\eqref{eq:RotaryPowerAGN} simplifies to
\begin{align}
\frac{\phi}{l} &= C_\phi \ \frac{\omega_p^2}{m_e^2} \ n_\nu \left(1 - r_\nu\right) \; .
\label{eq:NoDirectionality}
\end{align}

In contrast to the jet plasma, the inner regions of AGN accretion disks around SMBHs can reach electron densities as high as $n_e \sim 10^{14}-10^{16}$ cm$^{-3}$~\cite{Garcia:2016wse,1994ApJ...436..599S}, with spectral analyses of Seyfert galaxies frequently favoring inner-disk surface densities above $10^{15}$ cm$^{-3}$~\cite{Jiang:2019ztr}. Such densities imply a larger plasma frequency $\omega_p$ and hence a stronger prefactor $\propto \omega_{p}^2$ in Eq.~\eqref{eq:NoDirectionality}. At the same time, photons cannot propagate in regions where $\omega<\omega_p$; in practice this constraint is most relevant at radio/mm frequencies, whereas at higher frequencies, including UV/X-ray photons typically satisfy $\omega\gg\omega_p$, even for the highest densities quoted above. These dense conditions are realized in nearby Seyfert II nuclei such as NGC 1068, which has been identified as a potential UHE neutrino emitter by the IceCube Collaboration, despite showing no significant gamma-ray emission~\cite{IceCube:2022der}. Furthermore, recent IceCube analyses find a population-level correlation between neutrinos and X-ray bright Seyfert and radio-quiet AGN~\cite{Abbasi:2025tas}. In these objects, X-rays arise from a compact corona above the inner disk where thermal photons undergo inverse-Compton scattering. Additionally, IXPE~\cite{galaxies12050054} polarimetry further constrains the coronal geometry and magnetic structure, offering complementary insight into the regions of potential neutrino production.

The dominant production channel of UHE neutrinos in these inner-disk regions is hadronuclear $pp$ interactions arising from collisions of accelerated protons with the dense thermal plasma. These interactions produce roughly equal numbers of charged pions, $\pi^+$ and $\pi^-$, which subsequently generate a neutrino flux with nearly equal contributions of neutrinos and antineutrinos. A small but nonzero neutrino--antineutrino asymmetry is generically expected, since $pp$ collisions are not perfectly charge symmetric due to valence-quark effects, yielding $\pi^+/\pi^- \gtrsim 1$. In addition, subdominant hadronic processes such as $p\gamma$ interactions are $\pi^+$-biased, leading to slightly different $\nu$ and $\bar{\nu}$ yields~\cite{Biehl:2016psj,Fiorillo:2024jqz}. Overall for neutrinos produced in accretion disk plasma it is expected a suppressed asymmetry $(1-r_\nu) \ll 1$, compared to the maximal asymmetry of the jet scenario (see Sec.~\ref{sec:UHEnu}). With regard to concrete experimental data, the IceCube Collaboration’s observation of NGC~1068 currently provides only a muon-neutrino flux measurement~\cite{IceCube:2022der}, rather than an all-flavor flux normalization as for TXS~0506+056~\cite{IceCube:2024fxo}. Consequently, without knowing the electron-neutrino flux we cannot provide an estimate of the birefringence effect for such source. A full numerical treatment of neutrino production in accretion-disk environments would be required to predict such fluxes, which is beyond the scope of this work.

Given the above, we can nevertheless provide an estimate of the induced rotation angle due to the C$\nu$B permeating the accretion disk plasma. The effective propagation length of X-ray photons through the disk environment can be assumed to be of the order of a few gravitational radii $r_g = GM/c^2$, where $G$ is Newton's gravitational constant and $c$ the speed of light in vacuum. Namely, for a fiducial choice of a SMBH with mass $M = 10^9 M_\odot$, with $M_\odot$ denoting one solar mass, we have $l \equiv r_g = 4 \times 10^{-5}~\mathrm{pc}$, which is clearly much smaller than the typical parsec- to kiloparsec-scale characteristic of blazar jet plasma structures. However, for a high density plasma with $n_e = 10^{15}~\mathrm{cm}^{-3}$, and using Eqs.~\eqref{eq:CnuBasymmetry} and~\eqref{eq:RotaryPowerAGNCnuB} for C$\nu$B, we obtain:
\begin{equation}
    \phi_{\text{d}} \sim 10^{-35}\,\mathrm{rad}\,.
    \label{eq:EstimateCnuBDisk}
\end{equation}
This value, to our knowledge, is the largest neutrino-induced polarization-rotation angle reported in the literature.

\section{Concluding remarks}
\label{sec:conclusion}

In this work, we have investigated the parity-violating birefringence induced by an asymmetric neutrino background within the plasma environment of AGN, specifically focusing on relativistic jets and accretion disks. Our analysis demonstrates that the macroscopic parity violation arising from the neutrino-antineutrino asymmetry induces a rotation of the linear polarization plane of EM radiation propagating through the medium. A key finding is the derivation of a \emph{directionality factor} that accounts for the relative bulk motion between the neutrino medium and the plasma. We have shown that this factor introduces an anomalous frequency dependence for the rotation angle, distinct from the standard $\propto \omega^{-2}$ behavior of Faraday rotation, thus providing a unique spectral signature for neutrino-induced optical activity. 

We applied this formalism to three astrophysical scenarios: \emph{i}) relativistic jets traversing the C$\nu$B, \emph{ii}) jets with an internal flux of UHE neutrinos, and \emph{iii}) high-density accretion disks permeated by the C$\nu$B. For \emph{i}) the polarization rotation angle was estimated to be $\phi^{\rm C\nu B}_{\text{j}} \sim 10^{-41}$ rad, at radio frequencies. We identified a resonance condition $\omega \simeq \omega_p$, which can in principle enhance the effect, although this regime is not realized in realistic AGN jets. In \emph{ii}) we showed that if $E_\nu^{0} \simeq m_\nu \omega/\omega_p$ lies within the neutrino energy band, the birefringence angle acquires an anomalous frequency dependence that is qualitatively distinct from standard Faraday rotation. Adopting source parameters inferred by IceCube for TXS~0506+056, we found that the resulting rotation angles are exceedingly small $\sim 10^{-49}$–$10^{-67}~\mathrm{rad}$, across radio-to-X-ray frequencies. This strong suppression is primarily due to the low UHE neutrino flux. The maximum value is obtained for the optimistic assumption that the flux leads to a neutrino number density comparable to the jet plasma density of $n_e \sim 10^{4}~\mathrm{cm}^{-3}$, leading to $\phi^{\rm UHE}_{\rm j} \sim 5 \times 10^{-44}~\mathrm{rad}$, at radio frequencies. Finally, in case \emph{iii}), where electron densities can reach $n_e \sim 10^{15}~\mathrm{cm}^{-3}$, we obtained the largest rotation angle $\phi_{\rm d} \sim 10^{-35}~\mathrm{rad}$, at X-ray frequencies. While these values are below the sensitivity of current polarimeters, our work establishes a robust theoretical framework for these effects. 

Overall, the identified spectral discriminants offer a novel, magnetic-field-independent probe of neutrino asymmetries and AGN plasma environments, which may become accessible with future improvements in polarimetric sensitivity and the detection of denser neutrino environments.

\begin{acknowledgments}
The work of H.B.C. is supported by Fundação para a Ciência e a Tecnologia (FCT, Portugal) through the Projects UID/00777/2025 (https://doi.org/10.54499/UID/00777/2025) and CERN/FIS-PAR/0019/2021. H.B.C. thanks the Institute of Physics, Silesian University (Opava), for hospitality and financial support during the final stage of this work. We acknowledge the support of Czech Science Foundation (GAČR 24-12702S).
\end{acknowledgments}

%


\end{document}